\documentclass[twocolumn,showpacs,preprintnumbers,amsmath,amssymb]{revtex4}

\usepackage[colorlinks=true, %
  urlcolor=blue,   
  citecolor=blue,  
  linkcolor=red,   
  bookmarksopen=true]{hyperref}

\usepackage{graphicx}
\usepackage{dcolumn}
\usepackage{bm}
\usepackage{rotating}
\usepackage{verbatim}
\usepackage{xspace}
\usepackage{multirow}

\newcommand{\bea}{\begin{eqnarray}}
\newcommand{\eea}{\end{eqnarray}}

\newcommand{\beq}{\begin{equation}}
\newcommand{\eeq}{\end{equation}}

\newcommand{\refeq}[1]{Eq.~\eqref{#1}}

\newcommand{\noast}{\vphantom{\ast}}
\newcommand{\nodag}{\vphantom{\dagger}}

\def\BR{\ensuremath{{\cal{B}}(\tau \to \mu \gamma)}\xspace}

\begin{document}
\preprint{CERN-PH-TH-2010-189}
\preprint{DO-TH 09/09}
\preprint{HIP-2009-17/TH}

\title{A large Muon Electric Dipole Moment from Flavor?}

\author{Gudrun Hiller}
\affiliation{CERN, Theory Division, CH-1211 Geneva 23, Switzerland }
 \affiliation{Institut f\"ur Physik, Technische Universit\"at Dortmund,
  D-44221 Dortmund, Germany}
 \author{Katri Huitu}
 \affiliation{Department of Physics, and Helsinki Institute of Physics,
FIN-00014 University of Helsinki, Finland}
\author{Jari Laamanen}
\affiliation{Theoretical High Energy Physics,
Radboud University Nijmegen, P.O. Box 9010,
NL-6500 GL Nijmegen, The Netherlands}
\author{Timo R\"uppell}
 \affiliation{Department of Physics, and Helsinki Institute of Physics,
FIN-00014 University of Helsinki, Finland}

\begin{abstract}
  We study the prospects and opportunities of a large muon
  electric dipole moment (EDM) of the order $(10^{-24}-10^{-22})\, \rm{e
    cm}$.  We investigate how natural such a value is within the general
  minimal supersymmetric extension of the Standard Model with CP
  violation from lepton flavor violation in view of the experimental constraints. In models with hybrid
  gauge-gravity mediated supersymmetry breaking a large muon EDM is
 indicative for the structure  of flavor breaking at the Planck scale, and points towards a high messenger scale.
    \end{abstract}

\maketitle

\section{\label{sec:introduction} Introduction}

CP violating phenomena receive strong and continuous interest because
they provide a gateway to physics beyond the Standard Model (SM).
While CP violation is observed in the quark sector and is currently
believed to be predominantly stemming from the
Kobayashi-Maskawa-mechanism, further searches are being pursued to
test this (SM-)picture where CP and flavor violation are intimately
linked.  No breakdown of CP symmetry has been seen so far among
leptons, however, there has been great progress made in neutrino
masses and mixing.

We consider here lepton electric dipole moments (EDMs)
${\cal{L}}_{\rm EDM} =d_l (-i/2) \bar \psi_l \sigma_{\mu \nu} \gamma_5 F^{\mu \nu}
\psi_l $ \cite{Pospelov:2005pr}, specifically for muons, as probes of
CP and lepton flavor violation.  Lepton EDMs in the SM appear first at
four loop order, and are tiny, {\it e.g.}, $d_e^{\rm SM } \leq
10^{-38} {\rm ecm}$ \cite{Pospelov:1991zt}.  The current experimental
limits are \cite{Amsler:2008zz,Bennett:2008dy}
\begin{eqnarray} \label{eq:edm-data}
 d_e &=& (6.9 \pm 7.4) \cdot 10^{-28} \, {\rm e cm}  , \nonumber \\
d_\mu& = &(-0.1 \pm 0.9) \cdot 10^{-19} \,  {\rm e cm} .
\end{eqnarray}
While the bound on the electron EDM severely constrains flavor blind CP phases,
lepton flavor violating couplings are subject to the constraints from the branching ratios
of rare lepton decays. At 90 \% C.L. \cite{Amsler:2008zz,Aubert:2009tk}  
\begin{eqnarray}
{\cal{B}}(\mu \to e \gamma)  &< &1.2 \cdot 10^{-11} ,  \nonumber \\
{\cal{B}}(\tau \to \mu \gamma) &<& 4.4 \cdot 10^{-8} , \nonumber \\
{\cal{B}}(\tau \to e \gamma) &<& 3.3 \cdot 10^{-8} . \label{eq:decay-data}
\end{eqnarray}
This situation can change significantly due to dedicated experimental
initiatives in addition to the upcoming direct searches up to multi-TeV energies
at the LHC:
The MEG collaboration expects a reach in the
$\mu \to e \gamma$ branching ratio of order $10^{-13}$ in the next few years 
\cite{Signorelli:2003vw};
The bounds on the radiative tau decays can be improved
at a possible future super flavor factory up to $2 \cdot 10^{-9}$ 
\cite{Bona:2007qt} (with $75 {\rm ab}^{-1}$); there is a
recent proposal to improve the current limit on $d_\mu$ by three  orders of magnitude, 
even as low as  $5 \cdot 10^{-25} \, {\rm e cm}$ \cite{Adelmann:2006ab}.

We ask here whether $d_\mu$ can be large, {\it i.e.}, not
many orders of magnitude below its current bound and  if
$d_\mu$ is large, what are the requirements and implications  for 
flavor physics?

We address these questions within the minimal supersymmetric Standard
Model (MSSM). New sources of lepton flavor violation arise from
supersymmetry (susy) breaking contributions causing intergenerational
slepton mixing, for earlier works, see {\it e.g.},
\cite{Feng:2001sq,Bartl:2003ju,Babu:2000dq},  and   \cite{Ibrahim:2001jz} for flavor non-universal  but diagonal effects. With current data, and depending on the mass scale of the susy
spectrum,   for sizeable muon EDMs some fine-tuning is involved.
We measure its amount  within the general MSSM considering scenarios
where CP violation is  genuinely linked to lepton flavor violation only.
In this framework, there is no model-independent connection between
hadronic and leptonic CP violation, and we do not impose the  CP constraints from the hadron sector. Specifically, nuclear effects in  the extraction of the electron EDM
from  the Thallium EDM, see,  {\it e.g.,}  \cite{Pilaftsis:2002fe,Pospelov:2005pr},
and the 2-loop effects from  \cite{Pilaftsis:2002fe} are not included here.
The direct link to weak CP violation highlights the importance of the muon EDM with respect to the
nucleon ones.

Models with gauge mediation being the dominant source of
susy breaking but with an additional contribution from Planck-scale
gravity have recently been studied for their non-minimal flavor
properties \cite{Feng:2007ke,Hiller:2008sv}, but also the vacuum
structure \cite{Lalak:2008bc}.  We work out the conditions for a large
muon EDM for such a realistic hybrid model and argue that an observation in the range of the anticipated reach could be explained with very specific solutions to the flavor problem only.

The plan of the paper is as follows:
In Section \ref{sec:susygeneral} we introduce the basic MSSM parameters
relevant for the evaluation of the leptonic EDMs and rare decays, and discuss the
relation between these observables. 
A detailed numerical study  in the general MSSM is presented
in Section \ref{sec:MSSM}, where we also access
the fine-tuning required for a large muon EDM. Models based on hybrid gauge-gravity mediation are investigated in Section \ref{sec:hybrid}.
More conventions and formulae are given in the appendix.

\section{The Muon EDM with flavor \label{sec:susygeneral}}

In case of flavor blind CP violation only the
muon EDM is constrained by the tight limit on the electron one, Eq.~(\ref{eq:edm-data})
to be below, at  90 \% C.L.,
\beq   \label{eq:linear}
d_\mu  \sim m_\mu/m_e d_e <3.9 \cdot 10^{-25} \rm{ecm} ,
\eeq
where $m_l$  denotes the lepton masses.
We are thus led to consider
CP violation in flavor violation to obtain larger values of $d_\mu$. 
For the purpose of this work we set  all CP phases not related to flavor to zero.
To ease the notation we use $d_l$  interchangeably for both the EDM and its magnitude throughout this work.

In Section \ref{sec:slepton} we introduce the susy slepton flavor sector and define
the requisite mass insertion parameters. Constraints
from rare decays of leptons Eq.~(\ref{eq:decay-data}) put constraints on the amount of flavor violation. In Section \ref{sec:mass-insertions} 
we discuss the interplay between the muon EDM and  the rare decays.
In Section \ref{sec:allofthem} we investigate the higgsino contributions.

\subsection{Slepton Flavor \label{sec:slepton}}

Genuine susy  flavor violation enters through the structure of the soft terms in generation space. This concerns the mass-squared matrix of the charged sleptons, which is given by
\begin{equation} \label{eq:Mltilde2}
M_{\tilde l}^2=\left(\begin{array}{cc}
M_{LL}^{2\nodag} & M_{LR}^{2\nodag} \\
M_{LR}^{2\dagger} & M_{RR}^{2\nodag} \end{array}\right),
\end{equation}
and which connects left-chiral $\tilde l_L$ and
right-chiral sleptons $\tilde l_R$ as   $\tilde l^*_N M_{NM}^2 \tilde l_M$, with the
chiral projectors $N,M=L,R$. The  $3\times 3$ sub-matrices read as
\begin{eqnarray}
\label{eq:MLL}
M_{LL,ij}^2&=&
M_{L,ij}^2 + (F,D-\mbox{terms}),\nonumber\\ \label{eq:MLR}
M_{LR,ij}^2&=&v_1 A_{E,ij}- \mu v_2 (Y_e)_{ij},\nonumber\\
\label{eq:MRR}
M_{RR,ij}^2&=&M_{E,ij}^2 +(F,D-\mbox{terms}) .  \label{eq:M2char}
\end{eqnarray}
Here, $i,j=1,2,3$ denote  generational indices and $Y_e$ is the Yukawa matrix of the charged leptons. The Higgs vacuum expectation values $v_{1,2}$ obey 
$v=\sqrt{v_1^2 +v_2^2} \sim 174$ GeV and $v_2/v_1 =\tan \beta$. The $\mu$ parameter denotes the Higgs mass term from the MSSM superpotential.

The sneutrino masses cause intergenerational mixing as well. In the presence of left-handed neutrinos only, the mass-squared matrix is written as
\begin{equation}
M_{\tilde\nu,ij}^2=
M_{L,ij}^2 +(F,D-\mbox{terms}).  \label{eq:M2neu}
\end{equation}
In $M^2_{RR,LL}$ and $M^2_{\tilde \nu}$ we did not spell out explicitly the  flavor diagonal $F$- and $D$-terms whose effect on the diagonal entries is suppressed as
$m_l^2/M_{L,E}^2$ and $v^2/M_{L,E}^2$, respectively.
In the full numerical analysis these terms are included.

To make contact with the low energy phenomenology, we need to 
evaluate the soft terms at the weak scale, $m_Z$.
Further, we go over from the flavor eigenstates to the basis where the leptons are mass eigenstates and the neutralino interactions are diagonal in generation space.
We denote the respective unitary matrix of slepton-type $A=L_L,\nu_L,E$ by $V_A$,
and the corresponding slepton soft terms by a tilde, $\widetilde M^2_{LR}=V_{L_L} M^2_{LR} V_E^\dagger$ and $\widetilde M_A^2=V_A M^2_A V_A^\dagger$.
{}From the  latter the mass insertions can be read off as 
\bea \label{eq:deltas}
\delta^{L_L E}_{ij} & = & (\widetilde M_{LR}^2)_{ij}/M^2_{\tilde A},  \nonumber \\
\delta^A_{ij, \, i \neq j } & = & (\widetilde M_A^2)_{ij}/M^2_{\tilde A}, 
\eea 
where we introduced an average slepton mass  $M_{\tilde A }$. The $\delta$ parameters
Eq.~(\ref{eq:deltas})  induce
flavor changing neutral currents (FCNCs) and if complex, CP violation.

\subsection{$d_\mu$ versus \BR}
\label{sec:mass-insertions}

To understand the interplay between a large muon EDM but small enough
lepton flavor violating branching ratios we employ the following approximations:
We restrict ourselves to stau-smuon flavor mixing only, and neglect all lepton masses except for the one of the tau; we set  to zero the flavor violating chirality-flipping
couplings which are contained in the $A_E$-terms of \refeq{eq:MLR}.
Furthermore -- note that this only matters for the rare decays -- we assume that the leading contributions are  due to photino exchange. We use the formulae from Ref.~\cite{Gabbiani:1996hi}, which are obtained in the
mass insertion approximation, {\it i.e.,} for perturbative deltas sufficiently smaller than one.

In this approximation, the leading contribution to $d_\mu$ 
can be obtained at one-loop from
\begin{figure}
  \centering
  \includegraphics[width=5.cm]{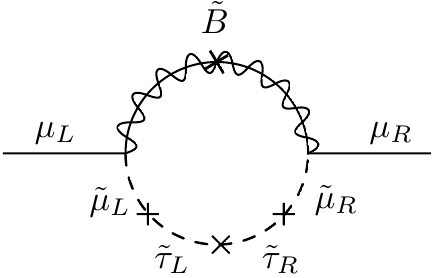}
  \caption{Flavor and chirality flow of the leading diagram contributing to the muon EDM. A cross denotes a mass insertion. The photon is attached wherever possible.}
  \label{fig:gr3}
\end{figure}
\begin{eqnarray} \label{eq:dmu-estimate}
\frac{d_\mu}{e} &= &\frac{\alpha}{2 \pi} \frac{M_i}{M_{\tilde A}^2} G_1(y) 
{\rm Im}( \delta^{L_L E}_{22}) ,
\end{eqnarray}
where $y=M_i^2/M_{\tilde A}^2$, and $M_i$ denotes the relevant gaugino mass, that is, here the
photino.
The loop function $G_1(y)$ obeys $G_1(1)=1/12$, drops rapidly with increasing $y$, and is given in the appendix. 

To estimate the EDM from flavor we approximate
$\delta^{L_L E}_{22}$ by  its effective value
$\delta^{L_L}_{23} \delta^{L_L E}_{33} \delta^{E *}_{23}$, see also Fig.~\ref{fig:gr3}, where the
bino contribution is shown.
Then, for $y=1$,
\bea
\label{eq:dmuapprox}
\frac{d_\mu}{e}  & \sim & 1 \cdot 10^{-20} {\rm cm} \left( \frac{200 \mbox{GeV}}{M_{\tilde A}} \right)   {\rm Im}
(\delta^{L_L}_{23} \delta^{L_L E}_{33} \delta^{E *}_{23} ) .
\end{eqnarray}

We allow for maximal CP phases of $\delta^{L_L}_{23} \delta^{E *}_{23}$. The factor $ \delta^{L_L E}_{33}$, see \refeq{eq:deltas},
parametrizes the left-right (LR) mixing of the staus, and is taken to be real here. In our analysis below we fix the value of $ \delta^{L_L E}_{33}$, thereby  linking the dependence on 
$A_{E, 33}, \mu, \tan \beta$ and $M_{\tilde A}$.

The $\tau \to \mu \gamma$ branching ratio can we written using the same approximations as
\bea \label{eq:tau-BR}
\BR = \kappa ( |\delta^{L_L}_{23}|^2 + |\delta^{E}_{23}|^2) ,
\eea
with
\bea \nonumber 
\kappa&=&  \left(1 +   \frac{M_i}{m_\tau} \frac{G_1(y)}{G_3(y)} \delta^{L_L E}_{33} \right)^2   \\
&\times & \frac{\alpha^3}{G_F^2} \frac{12 \pi}{M_{\tilde A}^4} G_3(y)^2   \times {\cal{B}}(\tau \to \mu \nu \bar \nu) . \label{eq:a}
\eea
We approximated
$\delta^{L_L E}_{23}$ and $\delta^{L_L E}_{32}$ by their effective values
$\delta^{L_L}_{23} \delta^{L_L E}_{33}$ and 
$\delta^{E}_{32} \delta^{L_L E}_{33}$, respectively. The loop function
$G_3(y)$ satisfies $G_3(1)=1/40$ and is given in the appendix.

The maximal value for the muon EDM allowed by the upper limit on the branching ratio 
$\BR_{\rm max}$ is then determined by \cite{Feng:2001sq}
\beq \label{eq:productmax}
\max|\delta^{L_L}_{23}  \delta^{E }_{23}| = \BR_{\rm max}/ (2 \kappa).
\eeq

\begin{figure}
\includegraphics[scale=0.8]{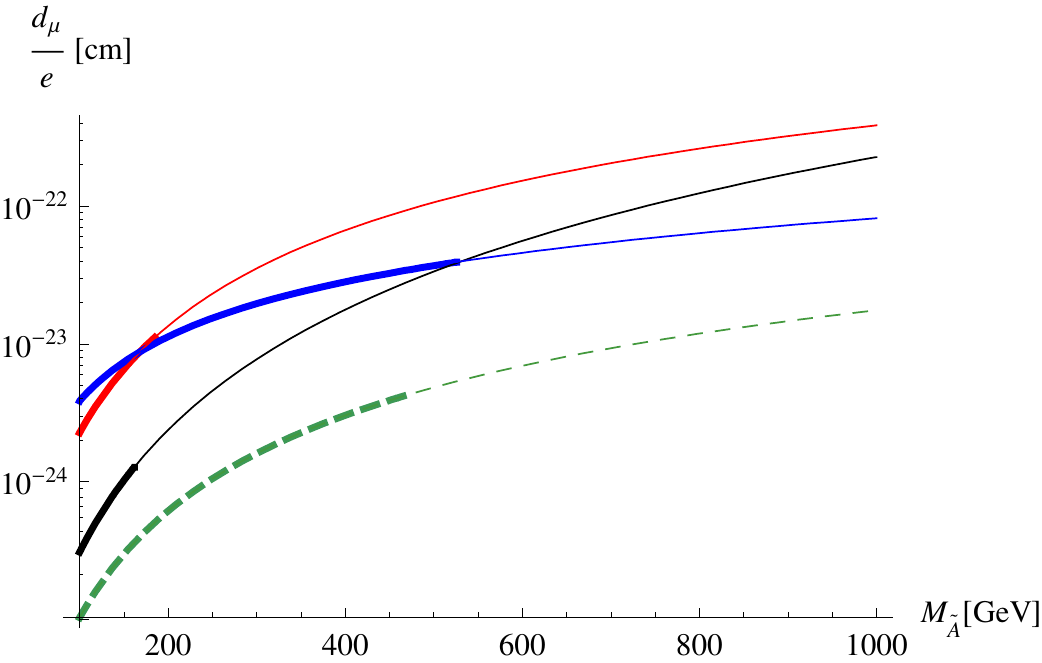}
\caption{\label{fig:dmumax}
The maximal value of  the muon EDM from flavor as a function of the average slepton mass 
based on
Eqs.~(\ref{eq:dmu-estimate})--(\ref{eq:productmax}), and $y=1$.
The three solid  curves  correspond to the current upper limit on \BR, \refeq{eq:decay-data},
and are obtained for $\delta^{L_L E}_{33}=10^{-3},10^{-4}$ and $10^{-2}$ (from top to bottom at
$M_{\tilde A}=1000$ GeV).
The dashed line refers to the hypothetical limit  \BR $<2 \cdot 10^{-9}$ for
$\delta^{L_L E}_{33}=10^{-3}$. The thick curves have mass insertions $|\delta| <1$.}
\end{figure}

\begin{figure}
\includegraphics[scale=0.8]{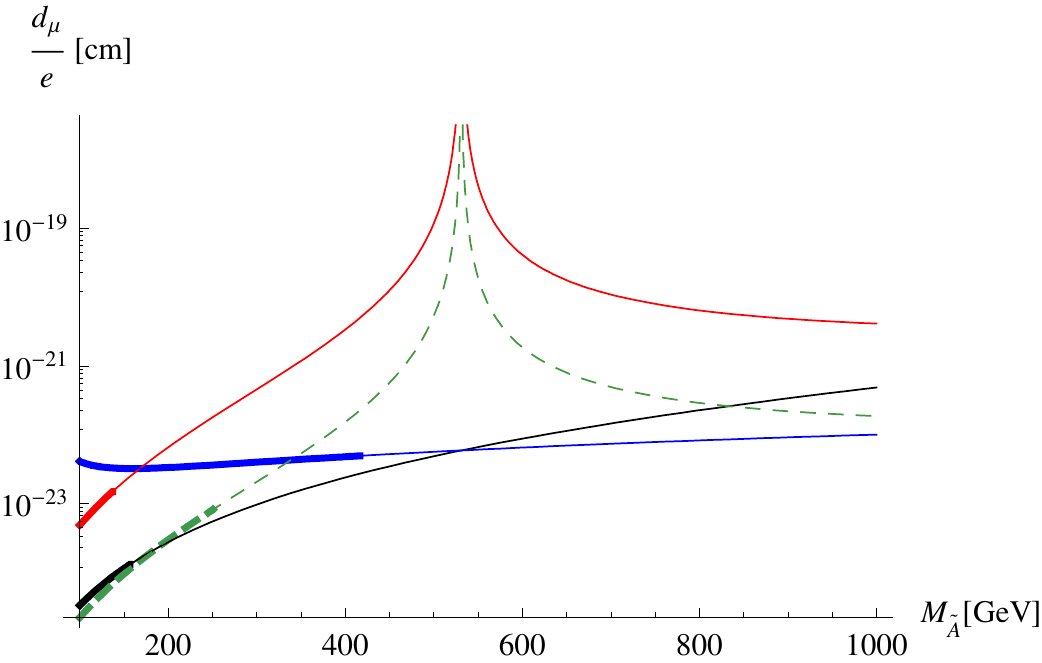}
\caption{\label{fig:dmumaxflip}
The same as in Fig.~\ref{fig:dmumax} but with the sign of $\delta^{L_L E}_{33}$ flipped, {\it i.e.}, being negative.}
\end{figure}

The resulting reach is shown in Figs.~\ref{fig:dmumax} and \ref{fig:dmumaxflip}.
The curves in these figures obtained
from extrapolations beyond the validity of the mass insertion approximation (thin lines)
are expected to illustrate the qualitative features only.
We used ${\cal{B}}(\tau \to \mu \nu \bar \nu) =17.36 \, \%$ \cite{Amsler:2008zz}.


We learn the following:
\begin{itemize}
\item Values of $d_\mu$ as large as ${\cal{O}}(10^{-22} ) \, \rm{ecm}$ are consistent with current FCNC constraints. For this to happen it requires   ${\cal{O}}(1)$ intergenerational mixing. (The mass insertion approximation breaks down).

\item The maximal allowed value of the muon EDM given by
\refeq{eq:productmax} grows with increasing $M_{\tilde A}$ and $y$. 

\item The muon EDM vanishes for vanishing  stau LR-mixing $ \delta^{L_L E}_{33}$ in our approximation. Large LR-mixing, however, suppresses $d_\mu$ because of the enhancement of the coefficient $\kappa$.
For our parameters we find  $\delta^{L_L E}_{33} \sim 10^{-(2-3)}$ to give the largest EDM, depending on the slepton mass.

\item The sign of the stau chirality mixing matters:  Negative values of $ \delta^{L_L E}_{33}$ 
allow  for larger EDMs by suppressing \BR, see \refeq{eq:a}, at the price of increased tuning.
This is illustrated in Fig.~\ref{fig:dmumaxflip}.

\item Within the mass insertion approximation, $|\delta| \lesssim 1$, values of $d_\mu$ 
up to  ${\cal{O}}(10^{-23}) \, \rm{ecm}$ are possible if the sleptons have masses below a few hundred GeV,  $ \delta^{L_L E}_{33}\sim 10^{-(2-3)}$ and $y \lesssim {\cal{O}}(1)$.

\item The maximal value for $d_\mu$ is proportional to the upper limit on \BR.
The anticipated future bound on \BR from super flavor factories of $2 \cdot 10^{-9}$ 
\cite{Bona:2007qt} has a significant impact on the maximal value of $d_\mu$.
\end{itemize}

\subsection{Including Higgsinos \label{sec:allofthem}}

Since we neglect the mass of the muon, the higgsinos (and winos) do
not contribute to $d_\mu$ to the order we are working, and
Fig.~\ref{fig:gr3} represents the leading contribution to the
EDM. However, higgsino contributions have been found to be of
importance in the calculation of \BR \cite{Feng:2001sq}. More
specifically, for a light $\mu$ term $\mu \sim M_1, M_2$ the lightest
neutralino has a substantial higgsino fraction and the photino-only
approximation in the rare $\tau$ decays \refeq{eq:tau-BR} receives
large corrections. Consequently, the bounds on flavor violation change
in the presence of the higgsinos, which then affects the maximal value
of $d_\mu$ from flavor.

\begin{figure}
  \includegraphics{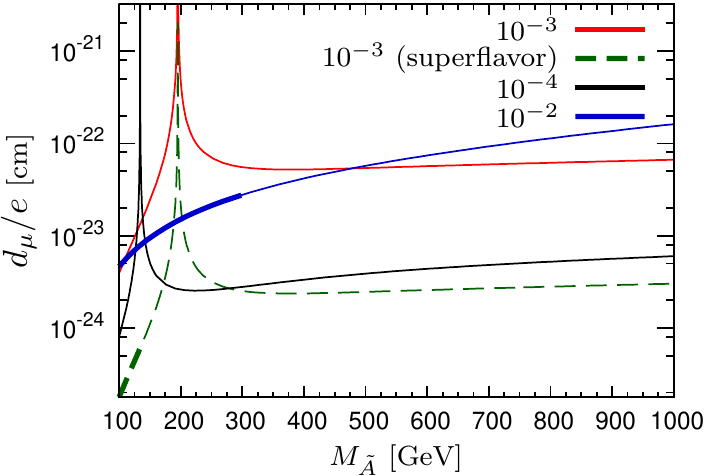}
  \caption{The maximal value of the muon EDM from flavor as a function
    of the average slepton mass based on \cite{Feng:2001sq} for $\tan
    \beta=3$, $M_1=M_{\tilde A}$ and $\mu=2$ TeV.  The three solid curves
    correspond to the current upper limit on \BR,
    \refeq{eq:decay-data}, and are obtained for $\delta^{L_L
    E}_{33}=10^{-2},10^{-3}$ and $10^{-4}$ (from top to bottom at
    $M_{\tilde A}=1000$ GeV).  The dashed line refers to the
    hypothetical limit \BR $<2 \cdot 10^{-9}$ for $\delta^{L_L
    E}_{33}=10^{-3}$. The thick curves have mass insertions $|\delta|
    <1$.}
  \label{fig:maxEdm}
\end{figure}
\begin{figure}
  \includegraphics{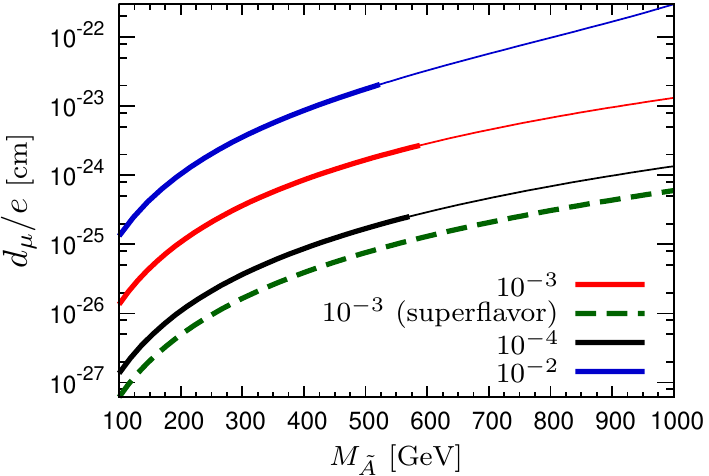}
  \caption{The same as in Fig.~\ref{fig:maxEdm} but with
    $\mu=M_{\tilde A}$, hence lighter and more important higgsinos.}
  \label{fig:maxEdmmumA}
\end{figure}

The higgsino contribution to the $\tau \to \mu \gamma$ amplitude
involves directly the $\tau$ Yukawa coupling $y_\tau \propto \tan
\beta$.  Hence, keeping $\delta^{L_L E}_{33}$ fixed while enhancing
$\tan \beta$, the higgsino loops get scaled up with respect to the
bino ones and the maximal value of the muon EDM drops with respect to
the analysis of the previous section.  The same holds if the $A$-term
dominates the stau LR-mixing, $\delta^{L_L E}_{33} \sim v A_{E,
33}/\tan \beta$.  If the $\mu$ term is large and dominates,
$\delta^{L_L E}_{33} \sim - \mu m_\tau \tan \beta$, then all bino and
higgsino contributions grow with $\tan \beta$. Since the EDM depends
only linearly on $\delta^{L_L E}_{33}$, whereas \BR does
quadratically, the maximal value of the EDM drops with increasing
$\tan \beta$ also in this case.  Therefore, large values of $d_\mu$
favor a low $\tan \beta$.

In Fig.~\ref{fig:maxEdm} the maximal muon EDM as a function of the
common slepton (smuon) mass is presented for the same values of
$\delta^{L_L E}_{33}$ as in Fig.~\ref{fig:dmumax}, however with the
higgsino and wino contributions in \BR included. We use $\tan\beta =
3,$ $\mu = 2$ TeV, $M_1=M_{\tilde A}$ and $M_2 = (g_2^2/g_1^2) M_1$.
As expected, a reasonably high value of $\mu$ suppresses the
contribution from the higgsino loops, and, in general, the increase in
\BR is small. Therefore, the photino analysis of
Sec.~\ref{sec:mass-insertions} gives a good approximation of the
maximum value of the muon EDM.
However, for low values of $\delta^{L_L E}_{33}$ the leading bino
graphs in the $\tau \to \mu \gamma$ amplitude, which are proportional
to $\delta^{L_L E}_{33}$, are suppressed, and external chirality flip
graphs $\propto m_\tau$, which are usually sub-leading, become more
important at the low slepton mass range. Therefore, a cancellation
against the higgsino graphs occurs, which can be seen in
Fig.~\ref{fig:maxEdm} as a spike in the allowed maximal EDM curves.  For
larger values of $\delta^{L_L E}_{33}$ the cancellation does not
happen, since the external flip contribution can not compete with the
leading contributions.

In Fig.~\ref{fig:maxEdmmumA} we show the maximal muon EDM for an
higgsino mass parameter $\mu$ which equals the slepton mass $M_{\tilde
A}$. Consequently, the smaller $\mu$ term enhances the higgsino
mediated loop contributions in \BR and the maximal EDM is not as large
as in Figs.~\ref{fig:dmumax} or \ref{fig:maxEdm}.  With increasing
slepton mass also $\mu$ increases, and the characteristics of
Figs.~\ref{fig:maxEdm} and \ref{fig:maxEdmmumA} become alike.

Here, we assumed that $\delta_{23}^{\nu_L}$ is equal to
$\delta_{23}^{L_L}$.  Typically the sneutrino loop gives the largest
single contribution to the branching ratio, unless the higgsinos are
decoupled, in which case the bino loops dominate.

To summarize, the inclusion of higgsino contributions modifies the
correlation between the maximal value of the muon EDM from flavor and
\BR given in the previous section for low values of the $\mu$ term or
when cancellations occur.  Values of $d_\mu$ at order $10^{-22} \,
\rm{ecm}$ are possible with current \BR constraints for order one
flavor mixings. Values of order $10^{-23} \, \rm{ecm}$ can be reached
within the mass insertion approximation, $|\delta| \lesssim 1$.

\section{Numerical analysis in the general MSSM \label{sec:MSSM}}


We study the muon EDM  in the general MSSM neglecting CP phases
unrelated to lepton flavor mixing.
Formulae used are given in the appendix.

\subsection{Input and Constraints on Susy Parameters}

We generate sets of data points by randomly sampling susy parameters
at the weak scale.  We consider two scenarios, a "light" and a "heavy"
one with their respective parameter ranges defined in Table
\ref{table:Input}.

\begin{table}[!h]
\begin{tabular}{r|c|c} 
& flavor diagonal 
& flavor off-diagonal \\ 
\hline
light & 0 -- 1 TeV & 0 -- 100 GeV \\ 
heavy & 3 -- 5 TeV & 0-- 3 TeV \\ 
\end{tabular}
\caption{\label{table:Input}The sampling ranges for the flavor   diagonal  and off-diagonal
mass parameters in the light and the heavy scenario. }
\end{table}

As can be seen from Table \ref{table:Input}, we use different ranges
for the flavor diagonal and off-diagonal mass parameters.  The former
contain the gaugino masses $M_1$ and $M_2$, the $\mu$ term, its susy
breaking companion, $B_\mu$, and the diagonal entries in the slepton
soft terms $\sqrt{M^2_{L, ii}}$, $\sqrt{M^2_{E, ii}}$, $A_{E, ii}$.
For simplicity we restrict our analysis to degenerate first and second
generation masses $M^2_{L(E), 11}=M^2_{L(E), 22}$. Diagonal $A$-terms
are assumed to follow the corresponding lepton Yukawa couplings $A_{E,
ii} =A_0 y_i$.  Furthermore, we include slepton flavor mixing between
the second and third generation only. The respective off-diagonal
parameters are therefore $A_{E, 23}, A_{E, 32}, \sqrt{M^2_{L, 23}}$
and $\sqrt{M^2_{E, 23}}$.  We allow for an order of magnitude
suppression of the off-diagonal soft terms with respect to the
diagonal entries in the light scenario, whereas in the heavy scenario
we allow the off-diagonal terms to be of the same order of magnitude
as the diagonal ones.  In both scenarios $\tan\beta$ is sampled over
the range 2 -- 50.

For each generated point we check the scalar potential against
unboundedness from below, and the existence of charge or color
breaking minima \cite{Casas:1996de}.  These bounds constrain the
off-diagonal soft contributions of the $A$-terms.  The soft masses of
the Higgs fields are solved for with the vacuum condition $\partial
V/\partial \phi = 0$.  The resulting spectrum at each generated point
is checked against experimental constraints from direct searches
\cite{Amsler:2008zz}.  To generate a muon EDM we introduce random CP
phases in all flavor off-diagonal soft terms between 0 and 2$\pi$.
Since we assume here the $\mu-e$ and $\tau-e$ slepton mixing terms and
all flavor blind CP violation to be zero, the experimental constraint
from the electron EDM is automatically fulfilled.  Contributions to
$\mu \to e \gamma$ decays from sneutrino loops are also suppressed and
neglected.

For each sample point, we calculate the contribution to the muon
anomalous magnetic moment, $\Delta a_\mu$, the muon electric dipole
moment $d_\mu$, and the branching ratio for the decay
$\tau\to\mu\gamma$. $\Delta a_\mu$ is constrained to be less
than $10^{-9}$ corresponding to roughly two sigma of the experimental
and theoretical uncertainty \cite{DeRafael:2008iu}. The 
bound on $\Delta a_\mu$  is always fulfilled in the heavy scenario.


\subsection{Random Walk Analysis}

\begin{figure}
\begin{tabular}{cccc}
\begin{sideways}\hspace{2cm}$\log(d_{\mu}[ \rm e cm ])$\end{sideways}&
\includegraphics[width=0.4\textwidth]{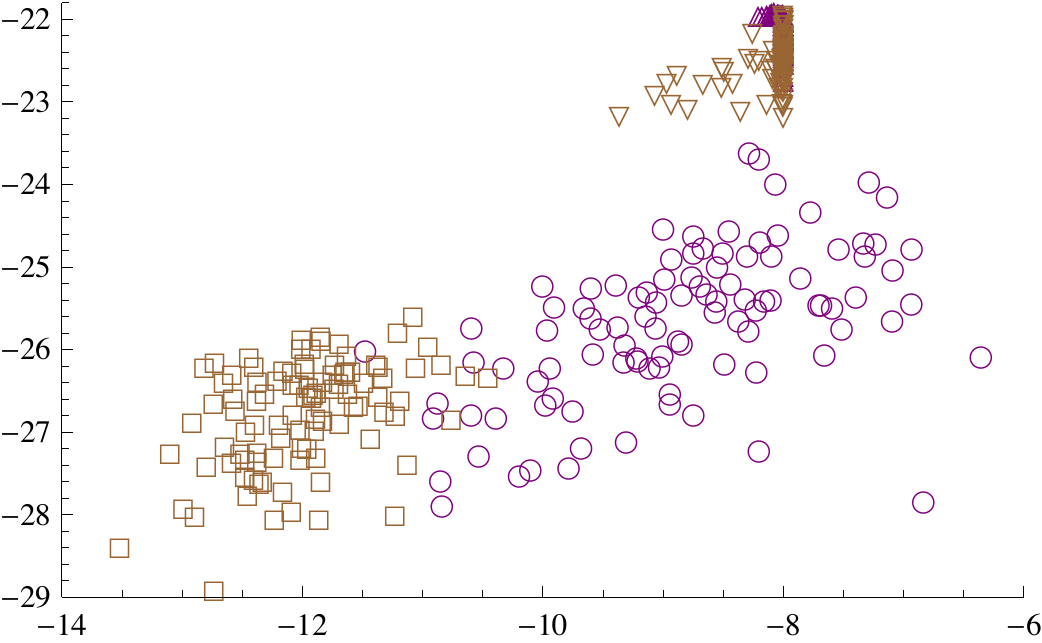} \\ &
\raisebox{0.1cm}[-0.1cm]{$\log(\BR)$} 
\end{tabular}
\caption{\label{plot:EDM-Br}The distribution of the muon EDM $d_{\mu}$
vs. the branching ratio \BR. Shown is the initial
distribution in the light
scenario (circles) and the heavy scenario (squares), as well as the
distribution after the random walk process in the light 
(triangles up) and the heavy scenario (triangles down).}
\end{figure}

The interplay between the muon EDM and the $\tau \to \mu \gamma$
branching ratio for our sampled points can be seen in
Fig.~\ref{plot:EDM-Br}. For both the light and heavy spectrum the
value of the EDM for points satisfying $\BR < 10^{-8}$ is typically
less than $10^{-24}$ ecm (squares and circles). This value is,
however, not a hard upper bound on the muon EDM, because it stems from
the stochastic nature of our parameter selection. Indeed, and as we
will see, one can find parameters such that the constraint on the
branching ratio is fulfilled and $d_\mu$ is large. The question then
is how fine-tuned such points are.

To obtain points with $d_\mu> (10^{-22} - 10^{-24})$
ecm obeying the $\tau \to \mu \gamma$ constraint we use a random walk process. Starting from
a point $p$ in parameter space, we randomly vary the parameters by a
small amount, $p \to p' = p+ \delta p$, and check for all the
constraints as well as for a larger value of the muon EDM.  If $p'$
passes the test, the procedure is repeated with $p'$ as the new
starting point. If we start out with a point $p$ that has
$\BR>10^{-8}$, we additionally require a decrease in the branching
ratio before accepting $p'$. The requirements on the branching ratio
and the muon EDM thus drive a point into the desired direction.
Unless we exceed a predetermined number of failed attempts, this
procedure eventually yields a point with $\BR< 10^{-8}$ and a large value of $d_\mu$ as requested, after which the random walk is terminated.

In Fig.~\ref{plot:EDM-Br} we show the result of this random walk in the heavy and the light
scenario.  We see a tendency of points that
start out with very low branching ratios to migrate to larger ones. This is expected, as the branching ratio has similar
contributions as the EDM, which is being forced to increasing values.
It should be noted that for the fixed number of iterations, we find
that in the heavy scenario only about 2\% of the original points
achieved $d_{\mu}=10^{-22}$ ecm, whereas in the light scenario this
fraction is  35\%. 

For each scenario we consider in the following  two data
sets, an inclusive one containing all the points
that went through the random walk routine and an exclusive one
containing only points that actually pass the EDM constraint
$d_{\mu} \ge 10^{-22}$ ecm for
the light scenario and $d_{\mu} \ge 0.5\cdot 10^{-22}$ ecm for the heavy
scenario.

\begin{figure}
\begin{tabular}{cccc}
\begin{sideways}\hspace{2cm}\# of points\end{sideways}&
\includegraphics[width=0.4\textwidth]{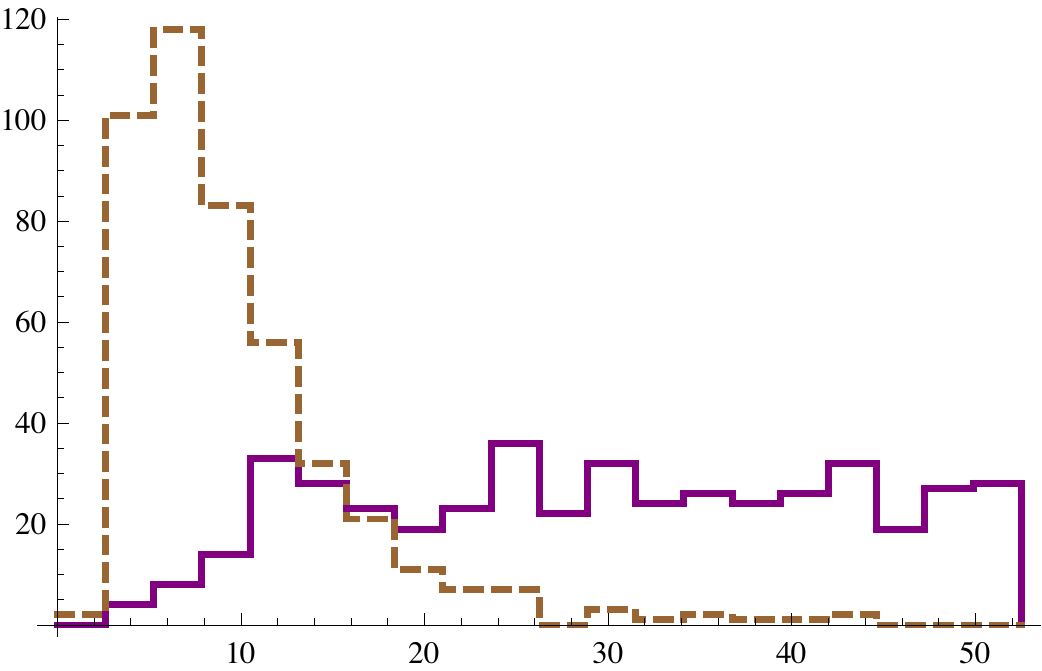} \\ &
\raisebox{0.1cm}[-0.1cm]{$\tan\beta$} 
\end{tabular}
\caption{\label{plot:tanBeta}The distribution of $\tan\beta$
before (solid line) and after (dashed line) the random walk process in
the light scenario (inclusive data set).}
\end{figure}

The random walk process affects the distribution of
the parameters. {\it E.g.}, in the light scenario there is a transition
from a near flat distribution of $\tan\beta$ to one which is peaked at
small values, see Fig.~\ref{plot:tanBeta}. The latter behavior is in agreement  with  the findings of
Sec.~\ref{sec:allofthem}.
In the next section we 
use this selection effect to construct a measure of fine-tuning.

\subsection{Measures of Fine-Tuning}
\label{sec:MeasuresFT}

In order to better understand the fine-tuning needed for achieving an
experimentally interesting $d_\mu$ while respecting the experimental
\BR constraint, we first note that the subamplitudes in
$a^{L/R}_{23}$, see Eq.~(\ref{eq:appaLR}), corresponding to neutralino
and chargino 1-loop Feynman diagrams, need to be small or cancel each
other to great accuracy. We find that, however, there is usually only
one subamplitude which contributes a dominant part of the branching
ratio. When we increase the EDM via our random walk routine this
disparity is lessened, but all individual subamplitudes remain small
enough for there to be no need for large cancellations.

Similarly, Eq.~(\ref{eq:edm}) suggests that the imaginary part of
$a^{R/L}_{22}$ needs to be as large as possible for a large EDM.
Since in our framework there is only an imaginary part in the flavor
off-diagonal 2-3 mixing, the subamplitudes in the initial data set
exhibit a dominance of the real part over the imaginary part for all
$a^{R/L}_{22}$'s.  A priori, one would expect this relation to be
reversed by the requirement to maximize $d_{\mu}$ but instead
we end up with a sample, where the
real and imaginary parts are within one order of magnitude of each
other in the light scenario, and only slightly more dispersed in the
heavy scenario. This is sufficient for generating a large EDM since
the overall sizes of the subamplitudes have increased. We conclude that
measures of fine-tuning which rely on looking at cancellations between
different subamplitudes in magnitude or in imaginary part are not
useful.

Another way of measuring fine-tuning is to look at the
local structure of parameter space and see how sensitive observables
are to variations of the parameters \cite{Barbieri:1987ed}. We find
that this measure of fine-tuning decreases for points when they go
through our random walk process. This is expected since this method of
measuring fine-tuning
effectively looks at the slope in parameter space of a given
observable, and in our case we drive these observables to their
extremum values, {\it i.e.}, to a region where the slope vanishes.

To discuss fine-tuning we instead consider the width of
a parameter's distribution and how the random walk process changes
that width (see Fig.~\ref{plot:tanBeta}). From several such changes in
the width we then construct a relative "volume" of parameter space
that is preferred by points that go through our random walk process.
We do this by comparing the number of fixed and equi-distant bins in a
parameter $X$ necessary to account for a given fraction $fr$ of points
before, $N_{\mathrm{before}}$, and after, $N_{\mathrm{after}}$, the
random walk process. This way, the bin number measures the width of a
(normalized) distribution. The ratio, $r_X
\equiv(N_{\mathrm{after}}/N_{\mathrm{before}})$, gives us a measure of
how much the range of the parameter $X$ has shrunk (peaked), $r_X<1$,
or expanded (flattened), $r_X>1$.  
Smaller (larger) volume fractions $r_X$  then correspond to
larger (smaller) tuning.
For the $\tan \beta$-distributions shown in 
Fig.~\ref{plot:tanBeta} we obtain $r_{\tan\beta}=0.25$ (for $fr=0.8$).

We consider the tuning of several parameters, specifically, the CP
phases of the four flavor off-diagonal soft terms, masses and $\tan \beta$.

\begin{figure}[ht]
\begin{tabular}{cc}
\begin{sideways}\hspace{1.5cm}\# of points\end{sideways}&
\includegraphics[width=0.4\textwidth]{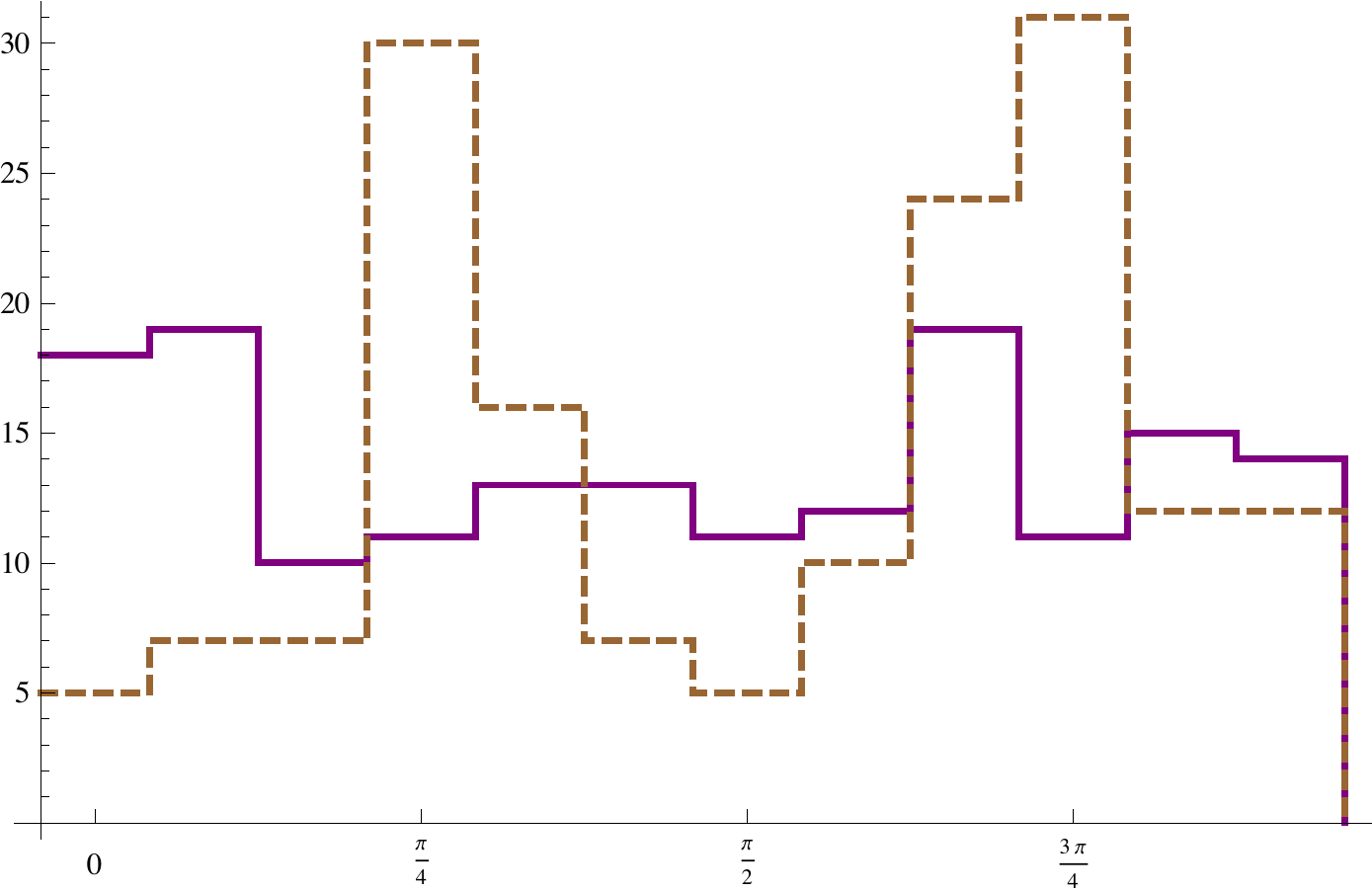} \\ &
\raisebox{0.1cm}[-0.1cm]{$\phi_{L_{23}}-\phi_{E_{23}}$}
\end{tabular}
\caption{\label{plot:difference}The distribution of
$\phi_{L_{23}}-\phi_{E_{23}}$ before (solid line) and after (dashed
line) the random walk process in the heavy scenario (exclusive data set).}
\end{figure}

Before proceeding, we note a couple of caveats. First, while random
walking the range of some parameters can expand beyond the range to
which we have initially constrained them, {\it e.g.}, the diagonal
soft terms in the heavy scenario are sampled between 3 and 5 TeV, and
the random walk process moves and spreads this range out.  Since this
is an artificial diffusion effect (and we can effectively choose $r_X$
to be whatever we want by adjusting the initial sampling range) we 
exclude such parameters directly from the measure of fine-tuning.
Secondly, even though a parameters distribution remains unchanged, it
may become correlated with another parameter.  An example for this
are the complex phases of the off-diagonal soft mass terms. Let $ 2
\phi_{X_{23}} = \arg (M^2_{X , 23})$ for $X=L,E$.  While we find for
each $\phi_{L_{23}}$ and $\phi_{E_{23}}$ separately the ratio $r\simeq
1$, their difference $\Delta \phi=\phi_{L_{23}}-\phi_{E_{23}}$ has $r_{\Delta \phi} \simeq 0.47$
with sharp peaks around $\pi/4$ and $3\pi/4$ (heavy scenario), as can be seen
in Fig.~\ref{plot:difference}. This correlation can be understood
from Eq.~\ref{eq:dmuapprox}, with
$ {\rm Im}(\delta^{L_L}_{23} \delta^{L_L E}_{33} \delta^{E *}_{23} )$ maximized.

We investigate the phase differences of $M^2_{X , 23}$, $X=L,E$ and
$A_{E, 23}, A_{E,32}$,
and assign the product of the individual volume changes $r_{\Delta \phi_i}$ as
\bea 
d_{\phi}=\prod_i  r_{\Delta \phi_i}.
\eea
Furthermore,  we consider various ratios of
mass parameters. We examine the following ratios containing
off-diagonal entries:
\beq
\label{eq:SlepPar}
\frac{|M^2_{L,23}|}{M^2_{L,33}},
\frac{|M^2_{E,23}|}{M^2_{E,33}},
\frac{|A_{E,23}|}{A_0},
\frac{|A_{E,32}|}{A_0}.
\eeq
The product of their individual volume fractions is denoted by $d_{od}$.
We look at some general ratios of parameters in the slepton sector
and the mass parameters $M_1$, $M_2$, $\mu$:
\bea
\label{eq:MiscPar} \nonumber
&&\frac{\sqrt{M^2_{L,22}-M^2_{L,33}}}{\frac{1}{2}(\sqrt{M^2_{L,22}}+\sqrt{M^2_{L,33}})}, 
\frac{\sqrt{M^2_{E,22}-M^2_{E,33}}}{\frac{1}{2}(\sqrt{M^2_{E,22}}+\sqrt{M^2_{E,33}})},  \\
&&\frac{\sqrt[4]{M^2_{L,33}M^2_{E,33}}}{A_0}, 
\frac{\sqrt[4]{M^2_{L,33}M^2_{E,33}}}{M_1}, 
 \frac{\sqrt{M_1 M_2}}{\mu},
\frac{M_1}{M_2}.
\eea
The product of their individual volume fractions is denoted by $d_G$.
We then define a total change of volume of parameter space, $d_V$, as
\bea \label{eq:dV}
d_V=d_\phi \cdot d_{od} \cdot d_G \cdot r_{\tan \beta} .
\eea

\begin{figure}[ht]
\begin{tabular}{cc}
\begin{sideways}\hspace{1.5cm}$\log (d_V)$\end{sideways}&
\includegraphics[width=0.4\textwidth]{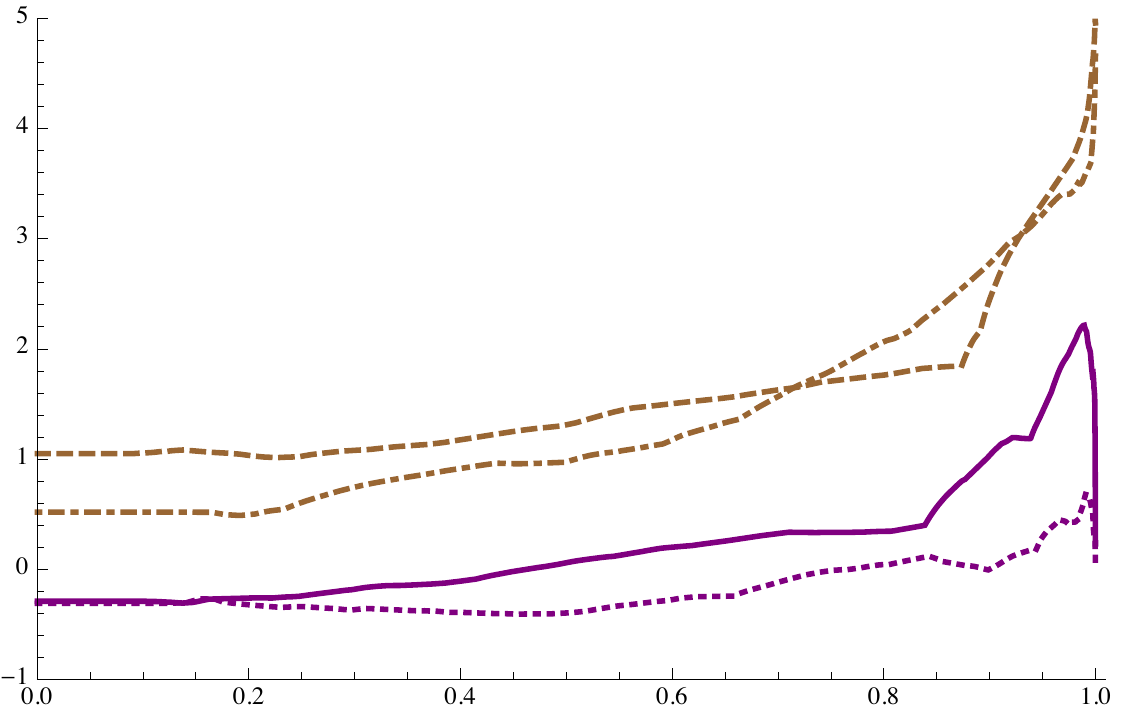} \\
& \raisebox{0.1cm}[-0.1cm]{$fr$}
\end{tabular}
\caption{\label{plot:pDependence}
The parameter space volume fraction $d_V$ Eq.~(\ref{eq:dV})
as a function of $fr$ in the light inclusive (solid line), light
exclusive (dotted line), heavy inclusive (dashed
line), and heavy exclusive (dash-dotted line) scenario.}
\end{figure}

One would expect that the volume changes $d_X$ depend on the value of the fraction of points, $fr$. 
In Fig.~\ref{plot:pDependence} we show this
dependence for $d_V$. As can be seen, $d_V$
is very stable over a large range of $fr$. The unstable behavior for 
values of $fr \gtrsim 0.8$ is due to the way statistical outliers affect
our analysis. A more sophisticated approach to measuring the width of
parameter distributions by fitting, {\it e.g.}, Gaussian curves could
improve the stability. We also verify that
the values of the $r_X$'s (and consequently $d_X$'s) remain stable
under variations of spurious quantities such as changes in data
binning.

\subsection{Discussion}
\label{sec:numericResults}

The various volume fractions for the two scenarios and data sets
are shown in Table~\ref{table:Pspace} for $fr=0.8$.
Comparing the different entries, we get an idea about 
fine-tuning caused by the random walk process.

Notably, the effect of the CP phases becoming
correlated (as shown in Fig.~\ref{plot:difference}) emerges  in
the exclusive  heavy scenario only. Also we see that a tuning of $\tan\beta$  occurs in the light scenario only (see Fig.~\ref{plot:tanBeta}).

\begin{table}[ht]
\begin{tabular}{|c|cc|cc|} 
\hline
&\multicolumn{2}{c|}{Inclusive}&\multicolumn{2}{c|}{Exclusive}\\
\cline{2-5}
& Light & Heavy & Light & Heavy \\
\hline
$r_{\tan\beta}$ & 0.25 & 0.96 & 0.16 & 0.95 \\
\hline
$d_{od}$ &  1.3 & 0.88 & 2.9 & 10 \\
$d_{\phi}$ & 0.98 & 0.96 & 0.98 & 0.49 \\
$d_G$ &  1.8 & 69 & 0.39 & 33 \\
\hline
$d_V$ &  2.2 & 59 & 1.1 & 170 \\
\hline
\end{tabular}
\caption{\label{table:Pspace}The volume fractions of parameter space
after the random walk for $fr=0.8$ for the light
and the heavy scenario. Smaller numbers correspond to larger tuning.}
\end{table}

The ratios of the off-diagonal to diagonal mass elements
which we have collected into $d_{od}$, have values close to unity in
the inclusive set but spread out, by up to an order of magnitude in
the case of heavy scenario, when we look at the exclusive set. This
means that the more fine-tuned the points get, the wider the range of
acceptable mass ratios is. These ratios enter  the loop
contributions of the branching ratio and EDM and their increase in
size, which leads to the spreading out of the distribution, is in line
with the observations from the beginning of Sec.~\ref{sec:MeasuresFT}.

Finally, we see that for the set of flavor diagonal mass
ratios in $d_G$, there is a fine-tuning effect in the light scenario
that emerges for the exclusive set, whereas in the heavy scenario there is
instead again a spreading out.

It is apparent that the ranges for the mass ratios are more
constrained in the light scenario. In the heavy scenario, the
individual loop contributions (which partly depend on these mass
ratios) start out much smaller than in the light scenario and thus
there is more ``room'' to spread out. The phases $\phi_{L_{23}}$
and $\phi_{E_{23}}$ becoming correlated is also a sign of 
insufficient size of the loop contributions.

It needs to be emphasized that the change in parameter space volume is a
relative measure of the amount of fine-tuning and useful for
comparisons between different data sets. It is not an absolute measure
of the ease of fine-tuning. Even comparing different
rows in Table~\ref{table:Pspace} is not trivial, {\it e.g.}, fine-tuning the
phases can have a greater or lesser effect on the EDM than fine-tuning
the mass ratios. 

We find that both heavy and light scenarios
appear to be not particularly tuned to give a large value of the muon EDM while respecting
existing constraints. While most of the volume
fractions $d_X$ in the light scenario are smaller than in the heavy one, from 
looking at Fig.~\ref{plot:EDM-Br} and noting
that a large percentage ($\sim$98\%) of points in the heavy
scenario can not be tuned to high enough values of $d_\mu$, we conclude that the heavy scenario is quantitatively harder to fine-tune.

\section{A Flavored EDM  in Hybrid models}
\label{sec:hybrid}

After analyzing the muon EDM within the general MSSM, we now consider the situation in
an explicit model with a hybrid mechanism of susy breaking, gauge-gravity mediation.
These models have recently received attention because they provide insights into the nature  of flavor breaking due to the appearance of flavor structures
beyond the Yukawa couplings \cite{Feng:2007ke,Hiller:2008sv,Nomura:2007ap,Hiller:2010dv}.
Viable gauge-gravity models have the dominant source of susy breaking from gauge mediation,
which is flavor-blind at the scale of mediation. Additional 
contributions arise from Planck-scale gravity, which generically affect flavor physics.
We assume here that the latter is controlled by Froggatt-Nielsen (FN) symmetries \cite{Froggatt:1978nt}, generating also the Yukawa matrices.

\begin{figure}[ht]
\includegraphics[width=0.4\textwidth]{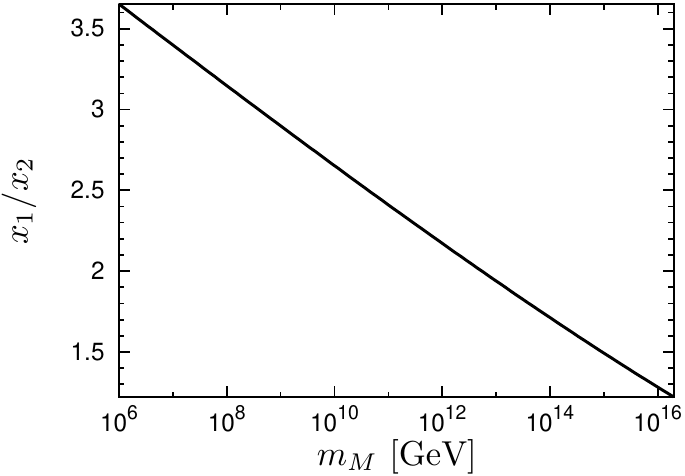} 
\caption{\label{fig:x1x2}The ratio $x_1/x_2$ in minimal gauge mediation Eq.~(\ref{eq:xi})
as a function of the messenger scale.}
\end{figure}

At the scale of gauge mediation, $m_M$, we parametrize the soft terms of the sleptons as
\cite{Feng:2007ke}
\begin{eqnarray}  \nonumber
M_L^2(m_M) &=& \tilde m_L^2 ( \mathbf{1}  + x_2 X_L ) , \\
M_E^2(m_M) &=& \tilde m_E^2 (\mathbf{1}  + x_1 X_R) .\label{eq:mLR2}
\end{eqnarray}
The new flavor structure from gravity with respect to gauge mediation is encoded
in the hermitean matrices $X_{L,R}$, following from the charge assignments of the
FN-symmetry. The factors $x_{1}$ and $x_2$ implicitly defined by Eq.~(\ref{eq:mLR2}) quantify
the relative size of the gravity versus the gauge contribution to the $SU(2)$ singlet  and doublet soft masses, respectively.
In this way, the $x_i$  are measures of the separation between the messenger and the Planck scale, 
$m_{\rm Pl}$. In minimal gauge mediation with $N_{M}$ messengers
(see,  {\it e.g.}, \cite{Martin:1997ns}),
\begin{equation} \label{eq:xi}
x_i 
\sim \left( \frac{m_M}{m_{\rm Pl}} \right)^2 \left( \frac{4
    \pi}{\alpha_i(m_M)} \right )^2  \frac{c_i}{N_M} , \, c_1=\frac{5}{6}, c_2=\frac{2}{3} ,
\end{equation}
where we assumed that the $F$-terms from $m_{\rm Pl}$ couple also to gauge mediation.
The ratio $x_1/x_2$ as in Eq.~(\ref{eq:xi}) is of order one, and increases for lower
values of the messenger scale, see Fig.~\ref{fig:x1x2}.

In Section \ref{sec:GG-deltas} we give formulae for the flavor violating low energy parameters in  gauge-gravity models. A flavor model with explicit FN charges is studied
in Section \ref{sec:modelB}. We investigate the impact of trilinear $A$-terms in
Section \ref{sec:Aterms}.

\subsection{Hybrid Gauge-Gravity \label{sec:GG-deltas}}

Starting from the soft slepton masses at the scale $m_M$
of the form given in Eq.~(\ref{eq:mLR2}) and solving the MSSM renormalization group (RG) equations 
\cite{Martin:1993zk}, the soft slepton masses at
the weak scale  can be written in the following approximate form:
\begin{eqnarray}
M^2_{L}(m_Z)& \sim&\tilde m^2_{L} (r_2 {\bf 1}+
c_L Y_e Y_e^\dagger +  x_2 X_{L}), \nonumber\\
M^2_{E}(m_Z)&\sim&\tilde m^2_E (r_1 {\bf 1}+
c_{E} Y_e^\dagger Y_e+ x_1 X_{R}) .  \label{eq:msoftmz} 
\end{eqnarray}
In the spirit of the FN flavor models, Eq.~(\ref{eq:msoftmz}) and the following
estimations are only accurate up to numbers of order one, which we indicate by "$\sim$".
The running of the gravity-induced terms is such an order one effect.
The coefficients $c_L,c_E$ are negative, obey $|c_L| <|c_E|$  and are of the order
$\sim 1/(16 \pi^2) \ln (m_M/m_Z)$, which are at most of order one for
$m_M$ near the GUT scale.

The $r_i$ coefficients include the leading RG correction to the flavor diagonal elements.
Neglecting contributions from
the sub-leading  gauge couplings, one obtains an analytical expression
for $r_i$ at one-loop order (see, {\it e.g.}, \cite{Martin:1997ns}):
\begin{equation} \label{eq:ri-exact}
r_i =1 + \frac{1}{c_i\pi} \left( \int_{\ln( m_Z)}^{\ln(m_M)}
dt  \frac{\alpha_i^3(t)}{\alpha_i^2(m_M)} \right)
\frac{M_i^2(m_M)}{ \tilde m_{i}^2(m_M)}  .
\end{equation}
Here, $M_i$ denotes the respective gaugino mass and $\tilde m_{i}^2$ 
equals $\tilde m_L^2$ for $i=2$ and $\tilde m_E^2$ for $i=1$. 
In messenger models of gauge mediation,
the ratio $M_i^2/\tilde m_{i}^2$ is determined by a simple formula at
the scale of mediation including only the leading gauge contributions:
\begin{equation} \label{eq:Mimtilde}
\frac{M_i^2(m_M)}{ \tilde m^2_{i}(m_M)}=c_i N_M .
\end{equation}
The RG parameters $r_i$ from Eqs.~(\ref{eq:ri-exact}) and (\ref{eq:Mimtilde}) obey 
$r_2 >r_1 >1$ and are 
of order one for not too many messengers.
Since $r_i >1$, 
the RG effect strengthens the flavor blind entries hence reduces the flavor violation.
The properties of the $r_i$ can be different in general gauge mediation 
\cite{Meade:2008wd} which does not exclude negative soft sfermion masses-squared at the mediation scale.

In the basis where the leptons are mass eigenstates
and the neutralino interactions are diagonal in generation space we obtain:
\begin{eqnarray}
\widetilde M^2_{L_L}(m_Z)& \sim&\tilde m^2_{L} (r_2 {\bf 1}+
c_L D_e^2+  x_2 V_{L_L} X_{L} V_{L_L}^\dagger ), \nonumber\\
\widetilde M^2_{\nu_L}(m_Z)& \sim&\tilde m^2_{L} (r_2 {\bf 1}+
c_L U^\dagger D_e^2U+  x_2 V_{\nu_L} X_{L} V_{\nu_L}^\dagger ), \nonumber\\
\widetilde M^2_{E}(m_Z)&\sim&\tilde m^2_E (r_1 {\bf 1}+
c_{E} D_e^2+ x_1 V_{E} X_{R} V_E^\dagger) , \label{eq:M2atmz}
\end{eqnarray}
where $U=V_{L_L} V^\dagger_{\nu_L}$ denotes the Maki-Nakagawa-Sakata (MNS) matrix and
$D_e={\rm diag}(y_e,y_\mu,y_\tau)=V_{L_L} Y_e V_{E}^\dagger$.
Note that the parametrization Eq.~(\ref{eq:M2atmz}) holds beyond minimal gauge mediation
with perturbative messengers.

The mass insertions  of the charged sleptons are then given as ($i \neq j$)
\begin{eqnarray}
\delta^{L_L}_{ij} \sim \frac{x_2}{r_2} (V_{L_L} X_{L} V_{L_L}^\dagger)_{ij} ,  ~~
\delta^{E}_{ij} \sim \frac{x_1}{r_1} (V_{E} X_{R} V_E^\dagger)_{ij} .
\end{eqnarray}
Here, we neglected contributions in the denominators from the tau-yukawa of the order
$y_\tau^2 \sim 10^{-4} \tan^2 \beta$ with respect to the $r_i$ for mixings involving the third generation.
The sneutrino mass insertions, which matter for the rare decays only to the order we are working,
$m_\mu=0$, are given as ($i \neq j$)
\begin{eqnarray} \label{eq:delnuL}
\delta^{\nu_L}_{ij} \sim \frac{1}{r_2} \left( c_L y_\tau^2 U^*_{3i} U_{3j}+  x_2 (V_{\nu_L} X_{L} V_{\nu_L}^\dagger)_{ij} \right) ,
\end{eqnarray}
receiving two independent competing contributions.
Note that the second term in Eq.~(\ref{eq:delnuL}) equals
$U^*_{ki} \delta^{L_L}_{kl} U_{lj}$. Since its sign/phase is not fixed by the FN symmetry, there can be cancellations in $\delta^{\nu_L}_{ij} $.

\subsection{A Flavor Model and Phenomenology \label{sec:modelB}}

Following  Ref.~\cite{Feng:2007ke} we entertain a $U(1) \times U(1)$ FN symmetry
with a single symmetry breaking parameter $\lambda$ of the order $0.1 -0.2$.
Specifically, we use  the following horizontal charges  
\begin{eqnarray} \label{eq:modelB}
L_1(2,0), L_2(0,2), L_3(0,2),  \nonumber \\ \bar E_1(2,1), \bar E_2(2,-1), \bar E_3(0,-1) ,
\end{eqnarray}
which result in a large 2-3 mixing.
The lepton mixing angles are obtained as
\bea
(V_{L_L})_{12} &\sim &\lambda^8  ,
(V_{L_L})_{13} \sim  \lambda^8  ,
(V_{L_L})_{23} \sim 1 , \nonumber \\
(V_{E})_{12} & \sim & \lambda^2 ,
(V_{E})_{13} \sim \lambda^4 ,
(V_{E})_{23} \sim \lambda^2 , \nonumber \\
(V_{\nu_L})_{ij} & \sim &1 ,
\eea
leading to $U_{ij} \sim {\cal{O}}(1)$. The current level of suppression of the 1-3 lepton mixing is considered accidental.

The slepton soft terms have the flavor structure
\bea
 (X_{L})_{12} & \sim & \lambda^4 ,
(X_{L})_{13}  \sim \lambda^4 ,
(X_{L})_{23} \sim 1 , \nonumber \\
 (X_{R})_{12} & \sim &  \lambda^2 ,
(X_{R})_{13} \sim \lambda^4 ,
(X_{R})_{23} \sim \lambda^2 ,
\eea
with the diagonal ones $(X_{L,R})_{ii} \sim 1$. Inserting the factors $X_{L,R}$ and the mixing angles into \refeq{eq:M2atmz} 
the flavor changing mass insertions are obtained as
\begin{eqnarray} \nonumber
\delta^{L_L}_{12} &\sim & \frac{x_2}{r_2} \lambda^4 ,  ~~
\delta^{E}_{12} \sim \frac{x_1}{r_1} \lambda^2 , ~~
\delta^{\nu_L}_{12} \sim \frac{1}{r_2} (c_L y_\tau^2 +x_2 )  ,\\
\delta^{L_L}_{13} &\sim & \frac{x_2}{r_2} \lambda^4 ,  ~~
\delta^{E}_{13} \sim \frac{x_1}{r_1} \lambda^4 , ~~
\delta^{\nu_L}_{13} \sim \frac{1}{r_2} (c_L y_\tau^2 +x_2)  , \nonumber \\
\delta^{L_L}_{23} &\sim &\frac{x_2}{r_2}  ,  ~~
\delta^{E}_{23} \sim \frac{x_1}{r_1} \lambda^2 , ~~
\delta^{\nu_L}_{23} \sim \frac{1}{r_2} (c_L y_\tau^2 +x_2) .
\end{eqnarray}
The relevant coupling for the muon EDM in this model is hence given as $\delta^{L_L}_{23} \delta^{E *}_{23}  \sim x_1  x_2/(r_1 r_2) \lambda^2 \lesssim 0.04$.

\begin{figure}
  \includegraphics{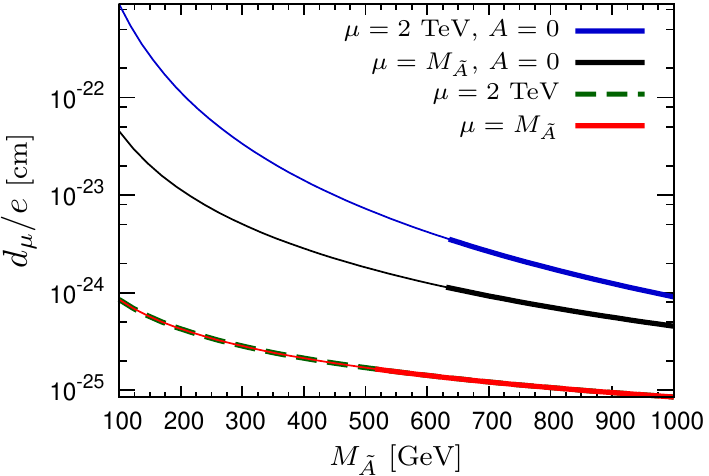}
  \caption{The muon EDM  in the flavor model Eq.~(\ref{eq:modelB}) with different 
  $\mu$ terms and stau LR-mixings. 
  Fad lines are consistent with the \BR bound \refeq{eq:decay-data}. In all curves
  $\tan \beta=3$ and $M_{\tilde A} =M_1$. The lower two, overlapping curves use
 $\delta^{L_L E}_{33}=10^{-3}$. }
    \label{fig:maxEdmGG}
\end{figure}

We illustrate the correlation between the rare decays and the muon EDM  for near maximal mass insertions $\delta^{L_L}_{23} = i 0.8$ and $\delta^{E }_{23} =0.05$ and various higgsino parameters and stau LR-mixings in Fig.~\ref{fig:maxEdmGG}.
Fad lines are in agreement with the current \BR bound. The two upper curves have a negatively
valued stau LR-mixing induced by the $\mu$ term. The lower two, overlapping curves
use $\delta^{L_L E}_{33}=10^{-3}$. Here, again, a larger $\mu$ term (dashed curve) gives heavier higgsinos and allows for a larger $d_\mu$  by relaxing the constraint from
the $\tau  \to \mu \gamma$ decays.

The constraint from the FCNC can be avoided in the hybrid models by suppressing flavor violation 
by increasing the separation between the Planck and the messenger scale.
The exact value of the upper bound on $x_i/r_i$ in the model Eq.~(\ref{eq:modelB})
depends on the susy spectrum and composition and is rather weak  due to the strong suppression of 1-2 mixing, which passes the $\mu \to e \gamma$ constraint.
For example, for the parameters along the fad lines in Fig.~\ref{fig:maxEdmGG} holds $x_i/r_i \lesssim {\cal{O}}(1)$.

We obtain in the flavor model Eq.~(\ref{eq:modelB}) for the maximal value of the muon EDM 
\bea \label{eq:FNupper}
\frac{d_\mu}{e}  
\lesssim   5 \cdot 10^{-24} {\rm cm} ,
\end{eqnarray}
in agreement with existing data, specifically the FCNC constraints \refeq{eq:decay-data}.
The other FN models of Ref.~\cite{Feng:2007ke}
have smaller values of $\delta^{L_L}_{23} \delta^{E *}_{23}$, hence yield smaller values of $d_\mu$, either due to a
smaller 2-3 mixing, or by a stronger constraint on $x_i/r_i$  effective for less suppressed 1-2 mixing. 

We conclude that an observation of the muon EDM at the level of order $10^{-(23-24)} \, {\rm e cm}$ would imply the following in the context of gauge-gravity models with a FN symmetry:
\begin{itemize}
\item The flavor structure needs to be very specific, {\it i.e.} not every FN symmetry correctly
reproducing the SM lepton masses and mixings works.
\item CP is broken together with flavor, and the CP phases are unsuppressed.
\item The susy spectrum is not too heavy, of the order of a few 100 GeV, and $y \lesssim 1$.
\item The messenger scale needs to be very high, not far from $m_M \sim \alpha/(4 \pi) m_{\rm Pl}$.
\end{itemize}

Note that analogous studies in  the quark sector have found that FCNC data 
constrain the messenger scale in typical FN  models to be about three orders of magnitude below $m_{\rm Pl}$  \cite{Hiller:2008sv}.
 
\subsection{Including $A$-terms \label{sec:Aterms}}

The gauge mediated contribution to the trilinear $A$-terms is of higher order and usually set to zero $A_E(m_M)=0$. This has also been employed implicitly in the previous section.
While the MSSM RG equations do induce finite $A$-terms at the weak scale, however, such terms
do not introduce CP or flavor violation beyond the Yukawa matrices, and do not matter for
flavor phenomenology. 

 However, in our  hybrid set-up, there can be a finite contribution from gravity to the $A$-terms
  \cite{Hiller:2010dv}:
\begin{equation} \label{eq:Ae-high}
A_E(m_M) \sim \sqrt{\tilde m^2 x} Y_e,
\end{equation}
where we used that the
gravity-induced $A$-terms receive the same parametric suppressions from the 
flavor symmetry breaking as the corresponding Yukawa matrices.
Here, $\tilde m^2, x$ are of the order of $ \tilde m_E^2$ and $\tilde m_L^2$,
$x_1$ and $ x_2$, respectively.

We then evolve according to the MSSM RG equations. The result can be recast in the
following approximate form
\begin{equation} \label{eq:Aelow}
A_E(m_Z) \sim Y_e ( a_E + \sqrt{\tilde m^2 x}+ b_E Y_e^\dagger Y_e ).
\end{equation}
The dominant contribution to $a_E$ and $b_E$ stems from the gaugino masses, which drives
them to magnitudes of electroweak size for large enough $m_M$. Note that
$a_E<0$ and $b_E>0$. Unless $\tan \beta$ is very large, the double Yukawa-suppression makes the $b_E$ term negligible.  As indicated in Eq.~(\ref{eq:Aelow}), the gravity contribution evolves with a coefficient of order one.

In the basis with lepton mass eigenstates and diagonal neutralino couplings we obtain:
\begin{eqnarray} \nonumber
\tilde A_E&=& V_{L_L} A_E V_E^\dagger \\
& \sim & D_e  ( a_E + \sqrt{\tilde m^2 x}   + b_E D_e^2 ). 
\end{eqnarray}

Under the RG evolution the $A$-terms mix onto the soft masses squared and induce also corrections to the soft masses in
Eq.~(\ref{eq:M2atmz}). Due to the Yukawa texture, this results effectively
into a correction of the order $x$ of the $c_{L,E}$ coefficients, and  will be absorbed therein for the purpose of this work.

\begin{figure}[ht]
\includegraphics[width=0.4\textwidth]{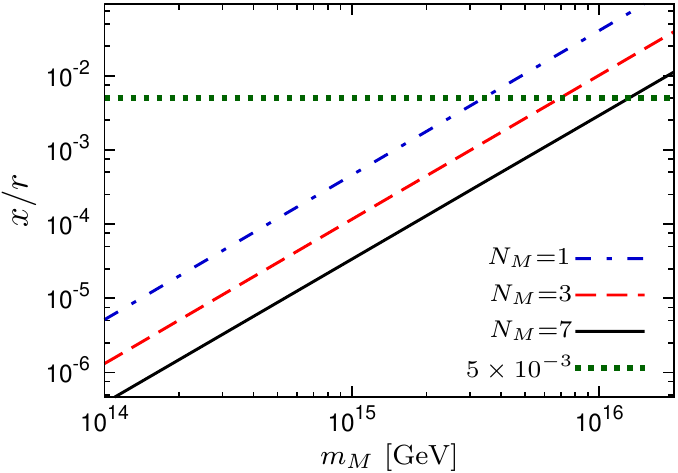} 
\caption{\label{fig:xoverr}The ratio $x/r$ in minimal gauge mediation Eq.~(\ref{eq:xi})
as a function of the messenger scale. The curves correspond to $N_M=1,3$ and 7 (from top to bottom). The horizontal line denotes the constraint from the
electron EDM, \refeq{eq:boundscale}.}
\end{figure}

The strongest bound on the hybrid model including $A$-terms Eq.~(\ref{eq:Ae-high}) is from the electron EDM, stemming from
$\delta^{L_L E}_{11}$ and its imaginary part. An approximate expression for $d_e$ reads as, see, \refeq{eq:dmu-estimate} for the corresponding formula for $d_\mu$,
\bea \nonumber
\frac{d_e}{e} &= &
\frac{\alpha}{2 \pi} \frac{M_i}{M_{\tilde A}^2} G_1(y) 
{\rm Im}( \delta^{L_L E}_{11}) , \\
& \lesssim & 2.6 \cdot 10^{-26} {\rm cm} \left( \frac{200 \mbox{GeV}}{M_{\tilde A}} \right)^2 \sqrt{ \frac{x}{r}},  \label{eq:e-edm}
\eea
where we assumed  $y=1$ in the second line.
Here, CP violation is induced by the gravity-mediated 
$A$-terms, Eq.~(\ref{eq:Ae-high}), which are fixed by the FN symmetry  up to order one, in general
complex numbers, only. 
One obtains from
Eqs.~(\ref{eq:edm-data}) and (\ref{eq:e-edm}):
\beq \label{eq:boundscale}
x/r \lesssim 5 \cdot 10^{-3}.
\eeq
This bound  implies a separation between the
Planck scale and the messenger scale of about up to four orders of magnitude, see Fig.~\ref{fig:xoverr}.
Information on the hadronic sector, that is, the neutron EDM in a similar set-up with
Yukawa-like $A$-terms and a viable quark FN symmetry gives somewhat milder constraints
 \cite{Hiller:2010dv}.

The bound Eq.~(\ref{eq:boundscale}) excludes a muon EDM from flavor since
the mass insertions are subsequently suppressed as
\beq
 \delta^{L_L}_{23} \delta^{E *}_{23}  \sim \lambda^k (x/r )^2 
\lesssim \lambda^k 10^{-5}  \leq 10^{-5},
\eeq
where $k$ denotes the non-negative integer power from the FN charges of a given model.

So far, we assumed the CP phase $\phi$ of $A_{E,11}$ to be fully unsuppressed.
The bound Eq.~(\ref{eq:boundscale}) weakens by a factor $1/\sin^2 \phi$ if $\phi$  is not maximal.
However, for $k=2$ as  in the  model from the previous section, even with a phase suppression as low as 
$\sin \phi$ at the percent level,  the prediction for $d_\mu$ is as in the linear mass 
scaling case, Eq.~(\ref{eq:linear}).

\section{\label{sec:summary} Summary}

Observing a large muon EDM  $d_\mu \gtrsim {\cal{O}}(10^{-24})\,  {\rm e cm}$  requires CP violation from flavor violation and favors a light mass
spectrum.  The heavier the spectrum and the larger the higgsino admixture, the larger is the requisite tuning to avoid the most relevant flavor constraint, \BR.

In the context of hybrid gauge-gravity models a measurement of such large values of the muon EDM would point towards
a specific Planck scale flavor structure and a very large  messenger scale not far from $\sim \alpha/(4 \pi) m_{\rm Pl}$. In the presence of Yukawa-textured $A$-terms at the level of Eq.~(\ref{eq:Ae-high})  with CP phases, however, 
the muon EDM
is limited by the electron EDM data to follow the linear mass scaling constraint \refeq{eq:linear}, which is below the current experimental sensitivity.

We conclude that the muon EDM is informative on various aspects of the underlying fundamental
physics.

\begin{acknowledgments}
KH and TR gratefully acknowledge the support from the Academy of
Finland (Project No. 115032). The work of JL is supported in part by the Foundation for Fundamental Research of Matter (FOM) and the Bundesministerium f\"ur Bildung und Forschung (BMBF).
\end{acknowledgments}

\appendix

\section{\label{sec:model} The relevant MSSM parameters}

The notation follows closely Ref.~\cite{Bartl:2003ju,Rosiek:1995kg}.

\subsection{Mixing matrices}
The slepton mass-squared  matrices Eqs.~(\ref{eq:Mltilde2})  and (\ref{eq:M2neu}) are diagonalized by the matrices 
$R^{\tilde l/\tilde\nu}$:
\bea
R^{\tilde l/\tilde\nu} M^2_{\tilde l/\tilde \nu} (R^{\tilde l/\tilde\nu})^{-1} 
&=& \mathrm{diag}(M^2_{\tilde l/\tilde \nu}).
\eea

The neutralino and chargino mass matrices are as usual
\bea
M_{\chi_0}&=&\left(\begin{array}{cccc}
M_1 & 0 & - {g' v_1 \over \sqrt{2}} & {g' v_2 \over \sqrt{2}} \\
0 & M_2 & {g v_1 \over \sqrt{2}} & - {g v_2 \over \sqrt{2}} \\
- {g' v_1 \over \sqrt{2}} & {g v_1 \over \sqrt{2}} & 0 & - \mu \\
{g' v_2 \over \sqrt{2}} & - {g v_2 \over \sqrt{2}} & - \mu & 0
\end{array}\right),\nonumber\\
M_{\chi_\pm}&=&\left(\begin{array}{cc}
M_2 & g v_2 \\
g v_1 & \mu \end{array}\right),
\eea
where $M_1$ and $M_2$ are the $U(1)$ and $SU(2)$ gaugino soft masses,
respectively, and $g' = e/\cos \theta_W = \sqrt{3/5}g_1$, $g =
e/\sin\theta_W =g_2$.  The neutralino and chargino mass matrices are
diagonalized by matrices $N$ and $U,V$ as follows:
\bea
U^{\ast} M_{\chi^\pm} V^{-1} &=& \mathrm{diag}(M_{\chi^\pm }) , \nonumber\\
N^{\ast} M_{\chi^0} N^{-1} &=& \mathrm{diag}(M_{\chi^0 }).
\eea

\subsection{Leptonic observables}

The relevant Lagrangian involving sleptons, neutralinos
and charginos is as follows:
\bea
\label{eq:appLvertex}
\mathcal{L}&=&\bar l_i(n_{ijk}^L L + n_{ijk}^R R)\chi_j^0 \tilde
l_k^{\noast}\nonumber\\
&&+\bar l_i(c_{ijk}^L L + c_{ijk}^R R)\chi_j^- \tilde \nu_k^{\noast}
+\mathrm{h.c.} ,
\eea
where
\bea
\label{eq:appcLR}
c_{ijk}^R &\equiv& -g V_{j1}^{\noast}R_{ki}^{\tilde\nu\ast},
\nonumber\\
c_{ijk}^L &\equiv& y_i^{\noast} U_{j2}^{\ast}R_{ki}^{\tilde\nu\ast} ,
\nonumber\\
\label{eq:appnL}
n_{ijk}^L &\equiv& -g'\sqrt{2}N_{j1}^{\ast}R_{k,i+3}^{\tilde l\ast}
-y_i^{\noast}N_{j3}^{\ast}R_{ki}^{\tilde l\ast},\\
\label{eq:appnR}
n_{ijk}^R &\equiv& \left(g' N_{j1}^{\noast}+g 
N_{j2}^{\noast}\right)R_{ki}^{\tilde l\ast}/\sqrt{2} 
-y_i^{\ast} N_{j3}^{\noast}R_{k,i+3}^{\tilde l\ast}. \nonumber
\eea
The susy contributions to the radiative lepton decay width, the lepton EDM, and the magnetic moment can be written at one-loop as \cite{Bartl:2003ju}
\begin{eqnarray}
  \label{eq:br}
  \Gamma (l_j \to l_i \gamma) &=& {\alpha m_{l,j}\over 16}(|a_{ij}^L|^2 +
  |a_{ij}^R|^2),\\
  \label{eq:edm}
  d_i &=& {e\over 4 m_{l,i}}\mathrm{Im}(-a_{ii}^L + a_{ii}^R),\\
  \Delta a_i &=& {1\over 2}\mathrm{Re}(a_{ii}^L + a_{ii}^R),
\end{eqnarray}
where terms of order $\mathcal O (m_{l_i} / m_{l_j})$ have been dropped. 
The coefficients
$a_{ij}^{L,R}$ are given as
\begin{eqnarray}
a_{ij}^L&=& {1\over 16 \pi^2}\sum_{k=1}^4 \sum_{r=1}^6 \left( \left(
n_{ikr}^L n_{jkr}^{L\ast} {m_{l,j}^2\over m_{\tilde\chi_k^0}^2}
+ n_{ikr}^R n_{jkr}^{R\ast} {m_{l,i}^2\over m_{\tilde\chi_k^0}^2}\right)
\right. \nonumber\\
&&\left. \times F_1\left({m_{\tilde l_r}^2 \over m_{\tilde\chi_k^0}^2}\right)
+ n_{ikr}^L n_{jkr}^{R\ast} {m_{l,j}\over m_{\tilde\chi_k^0}} 
F_3\left({m_{\tilde l_r}^2 \over m_{\tilde\chi_k^0}^2}\right)
\right)\nonumber\\
&+&{1\over 16 \pi^2}\sum_{k=1}^2 \sum_{r=1}^3 \left( \left(
c_{ikr}^L c_{jkr}^{L\ast} {m_{l,j}^2\over m_{\tilde\chi_k^+}^2}
+ c_{ikr}^R c_{jkr}^{R\ast} {m_{l,i}^2\over
m_{\tilde\chi_k^+}^2}
\right)\right. 
\nonumber\\
&&\left.\times F_2\left({m_{\tilde\nu_r}^2 \over m_{\tilde\chi_k^+}^2}\right)
+ c_{ikr}^L c_{jkr}^{R\ast} {m_{l,j}\over m_{\tilde\chi_k^+}} 
F_4\left({m_{\tilde\nu_r}^2 \over
m_{\tilde\chi_k^+}^2}\right)\right),\nonumber\\
a_{ij}^R&=&a_{ij}^L(L\leftrightarrow R),
\label{eq:appaLR}
\end{eqnarray}
where
\begin{eqnarray}
F_1(x) &=& -{2 + 3x - 6x^2 + x^3 + 6x\mathrm{log}x \over 6(1-x)^4},\label{app:Floop1}\\
F_2(x) &=& {1 - 6x + 3x^2 + 2x^3 - 6x^2\mathrm{log}x \over 6(1-x)^4},\label{app:Floop2}\\
F_3(x) &=& -{1 - x^2 + 2x\mathrm{log}x \over (1-x)^3},\label{app:Floop3}\\
F_4(x) &=& {1 - 4x + 3x^2 - 2x^2\mathrm{log}x \over (1-x)^3}.
\label{app:Floop4}
\end{eqnarray}

The  functions $G_{1,3}$ relevant for the calculation in  the mass insertion approximation are obtained from Eqs.~(\ref{app:Floop1}) and  (\ref{app:Floop3}) as $G_3(1/x) =(x^2/2) \cdot \partial F_1/\partial x$ and
$G_1(1/x) =(x^2/2) \cdot \partial F_3/\partial x$.

\end{document}